# TRIANGLES, TRIANGLES, TRIANGLES: CP VIOLATION IN THE B SYSTEMAND THE MEASUREMENT OF STRONG AND WEAK PHASES[1]

OSCAR F. HERNÁNDEZ[2] and DAVID LONDON[3]
*Laboratoire de Physique Nucléaire, Université de Montréal*
*Montréal, PQ, Canada H3C 3J7*

## ABSTRACT

We present the current wisdom regarding the measurement of the CP-violating phases of the CKM unitarity triangle in $B$-meson decays. After an introduction to the SM picture of CP violation, we review direct and indirect CP violation, the role of penguins and isospin analysis, and $B \to DK$ decays. We also discuss recent work on how to use SU(3) flavor symmetry, along with some dynamical approximations, to get at the CKM weak phases. Through time-independent $B$-decay measurements alone, we show that it is possible to extract all information: the weak phases, the incalculable strong phase shifts, and the sizes of the tree, color-suppressed, and penguin contributions to these decays.

## 1. Introduction

### 1.1. The Unitarity Triangle

While macroscopic physical phenomena appear to be invariant under the charge-conjugation–parity (CP) discrete symmetry of the Poincaré group, the discovery of the decay $K_L \to \pi\pi$ decay showed that the fundamental microscopic laws violate CP. In the near future, the study of $B$-meson decays will be a crucial testing ground for the Standard Model (SM) picture of CP violation, which is based on phases in the Cabibbo-Kobayashi-Maskawa (CKM) matrix [1]. In studying CP violation in the $B$ system, it is convenient to use an approximate form of the CKM matrix, due to Wolfenstein [2]. This approximation is based on the observation that the elements of

---

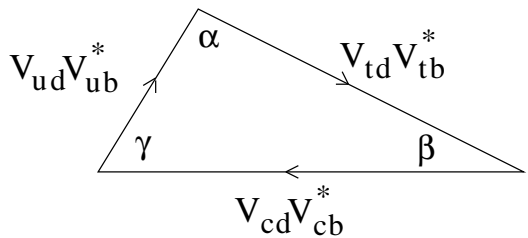

Figure 1: The unitarity triangle.

the CKM matrix obey a hierarchy in powers of the Cabibbo angle, $\lambda = 0.22$:

$$\begin{pmatrix} V_{ud} & V_{us} & V_{ub} \\ V_{cd} & V_{cs} & V_{cb} \\ V_{td} & V_{ts} & V_{tb} \end{pmatrix} \sim \begin{pmatrix} 1 - \frac{1}{2}\lambda^2 & \lambda & |V_{ub}|\exp(-i\gamma) \\ -\lambda & 1 - \frac{1}{2}\lambda^2 & A\lambda^2 \\ |V_{td}|\exp(-i\beta) & -A\lambda^2 & 1 \end{pmatrix} . \quad (1)$$

Here, $A$ is a parameter of $O(1)$, and $|V_{ub}|$ and $|V_{td}|$ are terms of order $\lambda^3$. In this approximation, the only non-negligible complex phases appear in the terms $V_{ub}$ and $V_{td}$. Unitarity of the CKM matrix implies, among other things, the orthogonality of the first and third columns:

$$V_{ud}V_{ub}^* + V_{cd}V_{cb}^* + V_{td}V_{tb}^* = 0 . \quad (2)$$

This relation can be represented as a triangle in the complex plane (the unitarity triangle), as shown in Fig. 1. In the Wolfenstein approximation, the angles in the unitarity triangle are given by $\beta = -\mathrm{Arg}(V_{td})$, $\gamma = \mathrm{Arg}(V_{ub}^*)$, and $\alpha = \pi - \beta - \gamma$ [3]. The SM picture of CP violation can thus be tested by independently measuring the three angles $\alpha$, $\beta$ and $\gamma$ and seeing (i) that they are all different from 0 or $\pi$, and (ii) that they add up to $\pi$ radians.

1.2. CP-violating asymmetries in B decays

The most promising signal for CP violation in the $B$ system is in rate asymmetries. That is, one looks for a difference between the rate for a $B$-meson to decay to a final state $f$ [$\Gamma(B \to f)$] and that of the CP-conjugate process [$\Gamma(\bar{B} \to \bar{f})$]. In order to produce such an asymmetry, it is necessary that (at least) two amplitudes with a relative phase contribute to the process $B \to f$. There are two distinct ways in which this can happen, called "direct" and "indirect" CP violation. We discuss these in turn.

• Direct CP Violation

Here the two amplitudes which interfere with each other contribute directly to the decay of the $B$ meson. For example, consider the decay $B^+ \to \pi^0 K^+$. In this case, there are contributions from both tree and penguin diagrams (Fig. 2). The phases associated with the two diagrams can be separated into two types: "weak" and "strong" phases. The weak phases $\phi$, which are due to the CKM matrix elements, change sign when one goes from $B^+ \to \pi^0 K^+$ to $B^- \to \pi^0 K^-$. On the other hand, the strong phases $\delta$, which are due to hadronization and final-state meson rescattering effects, are the same for both the decay and the CP-conjugate decay. This is due to the fact that the strong interactions are sensitive to color only – it is irrelevant whether a quark or an anti-quark is involved. Thus, the amplitudes for this decay and its CP-conjugate can be written as

$$\begin{aligned}
A(B^+ \to \pi^0 K^+) &= T\,e^{i\phi_T}\,e^{i\delta_T} + P\,e^{i\phi_P}\,e^{i\delta_P}\,, \\
A(B^- \to \pi^0 K^-) &= T\,e^{-i\phi_T}\,e^{i\delta_T} + P\,e^{-i\phi_P}\,e^{i\delta_P}\,.
\end{aligned} \quad (3)$$

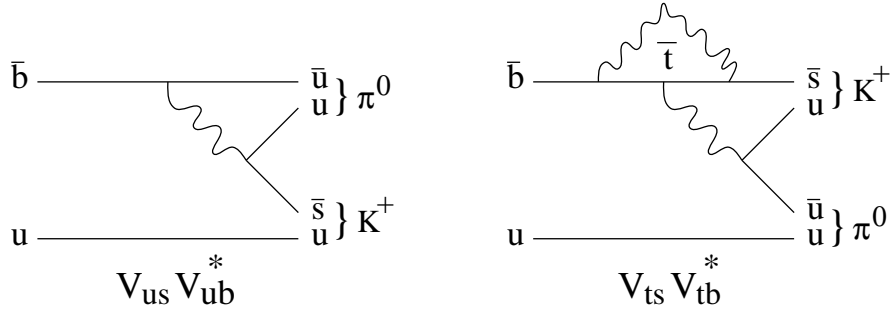

Figure 2: Two diagrams contributing to the process $B^+ \to \pi^0 K^+$.

(In this case $\phi_T = arg(V_{ub}^* V_{us}) = \gamma$ and $\phi_P = arg(V_{tb}^* V_{ts}) = \pi$.) It is straightforward to show that the difference in the decay rates is

$$\Gamma(B^+ \to \pi^0 K^+) - \Gamma(B^- \to \pi^0 K^-) \sim \sin(\phi_T - \phi_P)\sin(\delta_T - \delta_P). \quad (4)$$

Note that, although this rate asymmetry is proportional to $\sin(\phi_T - \phi_P) \sim \sin\gamma$, it also depends on the strong phase difference $\sin(\delta_T - \delta_P)$. The problem is that these strong phases are incalculable. Thus, a measurement of the rate asymmetry in $B^+ \to \pi^0 K^+$ does not provide *clean* information on the CKM phases. This is true of all processes which involve direct CP violation. (As we will see later, however, there are techniques to separate the weak and the strong phases, so that direct CP-violation measurements can in fact be used to extract the weak phases cleanly. In this particular example, SU(3) flavor symmetry can be used.)

• *Indirect (Mixing-Induced) CP Violation*

Indirect CP violation is due to $B^0$-$\overline{B^0}$ mixing. One chooses a final state $f$ to which both $B^0$ and $\overline{B^0}$ can decay. In this case it is the interference between the two amplitudes $B^0 \to f$ and $B^0 \to \overline{B^0} \to f$ which gives rise to CP violation. In order to be able to obtain clean CKM phase information, it is a necessary requirement that only one weak amplitude contribute to the decay. If more than one amplitude contributes, then direct CP violation is introduced, ruining the cleanliness of the measurement.

If the final state $f$ is a CP eigenstate, then clean information about CKM phases can be extracted from the time-dependent rates $B^0(t) \to f$ and $\overline{B^0}(t) \to f$. Here, $B^0(t)$ [$\overline{B^0}(t)$] is a state which is produced as a $B^0$ [$\overline{B^0}$] at time $t = 0$. Due to $B^0$-$\overline{B^0}$ mixing it will evolve in time into a mixture of $B^0$ and $\overline{B^0}$. If $f$ is not a CP eigenstate, it is necessary to measure the four rates $B^0(t) \to f$, $\overline{B^0}(t) \to f$, $B^0(t) \to \bar{f}$ and $\overline{B^0}(t) \to \bar{f}$ [4].

There are three classes of such CP asymmetries which are expected to be nonzero in the SM [3]. Along with the CKM angles which they measure, these are:

1. $\overset{(-)}{B_d}$ decays with $b \to u$ (e.g. $\overset{(-)}{B_d} \to \pi^+\pi^-$): $\sin 2\alpha$
   [WARNING: possible penguin hazard – see below]

2. $\overset{(-)}{B_d}$ decays with $b \to c$ (e.g. $\overset{(-)}{B_d} \to \Psi K_S$): $\sin 2\beta$

3. $\overset{(-)}{B_s}$ decays with $b \to u$ (e.g. $\overset{(-)}{B_s} \to D_s^+ K^-$, $D_s^- K^+$ [5]): $\sin^2 \gamma$

(One can also look for CP asymmetries in $\overset{(-)}{B_s}$ decays with $b \to c$ (e.g. $\overset{(-)}{B_s} \to \Psi \phi$), but these are expected to be very small in the SM.) It is therefore possible to cleanly extract the three angles of the unitarity triangle ($\alpha$, $\beta$, $\gamma$) using decays in which mixing-induced CP violation occurs, and indeed future experiments will use such processes to search for CP violation. We emphasize, however, that such measurements require (i) time-dependent information, and (ii) tagging, i.e. the knowledge of whether the decaying neutral $B$ meson was a $B^0$ or a $\overline{B^0}$ at time $t = 0$.

*1.3. Penguin Pollution and Isospin*

In the previous subsection we noted that, if one hopes to cleanly extract CKM phases using indirect CP violation, it is important that only one weak amplitude contribute to the decay. However, a problem arises when one realizes that, in fact, many $B$ decays have more than one such contribution. In particular, in addition to tree diagrams, penguin diagrams are often present [6, 7]. This is the case, for example, in the decay $\overset{(-)}{B_d} \to \pi^+\pi^-$, as shown in Fig. 3. Here, the tree diagram has the weak phase $V_{ub}^* V_{ud}$ ($\sim \gamma$), while that of the penguin diagram is $V_{tb}^* V_{td}$ ($\sim \beta$). In other words, in this decay, in addition to indirect CP violation, direct CP violation is present due to the interference of the tree and penguin diagrams. As discussed above, the presence of direct CP violation spoils the cleanliness of the measurement, hence the term "penguin pollution." Thus, a measurement of the CP asymmetry in this mode does not give $\sin 2\alpha$, as advertised above, but rather $\sin(2\alpha + \theta_{+-})$, where $\theta_{+-}$ depends on the weak and strong phases of the tree and penguin diagrams, as well as on their relative sizes. (We note in passing that the same problem does not arise for the decays $\overset{(-)}{B_d} \to \Psi K_S$ and $\overset{(-)}{B_s} \to D_s^+ K^-$, $D_s^- K^+$. In the former case, both penguin and tree diagrams have the same weak phase, so there is no interference, and in the latter case there are no contributions from penguin diagrams.)

All is not lost, however. Even in the presence of penguin diagrams, it is still possible to cleanly extract the CKM phase $\alpha$ by using an isospin analysis [8]. The idea is to use isospin to relate the three amplitudes $A(B_d^0 \to \pi^+\pi^-)$, $A(B_d^0 \to \pi^0\pi^0)$ and $A(B^+ \to \pi^+\pi^0)$, and similarly for the CP-conjugate processes. For all these decays, the final state has total isospin $I = 0$ or $2$. In other words, there are two amplitudes for these decays: $\Delta I = 1/2$ and $\Delta I = 3/2$. Since there are two isospin amplitudes, but three $B$-decay amplitudes, there must be a triangle relation among the $B$ amplitudes. It is:

$$\frac{1}{\sqrt{2}} A^{+-} + A^{00} = A^{+0} \ . \tag{5}$$

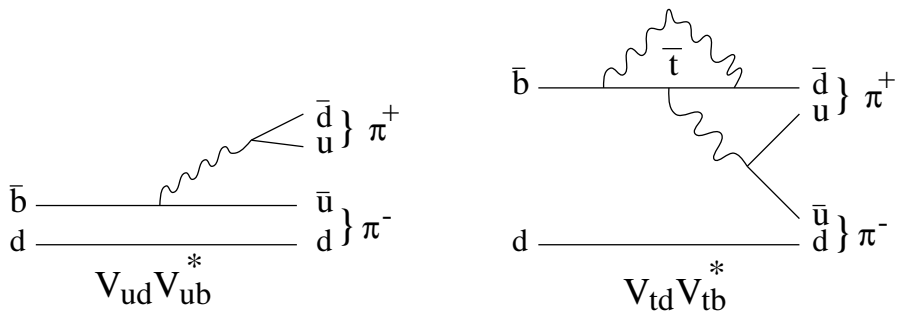

Figure 3: Diagrams contributing to the process $B_d^0 \to \pi^+\pi^-$.

There is a similar relation among the CP-conjugate processes:

$$\frac{1}{\sqrt{2}} \bar{A}^{+-} + \bar{A}^{00} = \bar{A}^{-0} \ . \tag{6}$$

Note that the tree diagram has both $\Delta I = 1/2$ and $\Delta I = 3/2$ pieces, but the penguin diagram is pure $\Delta I = 1/2$. Thus, if we can isolate the $\Delta I = 3/2$ contribution, we will have removed the "penguin pollution."

In fact, it is possible to isolate the $\Delta I = 3/2$ contribution by measuring $\theta_{+-}$ experimentally. This is done using the above triangle relations. The key point is that the decay $B_d^0 \to \pi^+\pi^0$ has no penguin contribution. Therefore there is only one weak amplitude for this decay, which means that there is no CP violation. In other words, $|A^{+0}| = |\bar{A}^{-0}|$, so that the two triangles have one side in common. (The other sides are not equal in magnitude, in general. Due to direct CP violation, we expect $|A^{+-}| \neq |\bar{A}^{+-}|$ and $|A^{00}| \neq |\bar{A}^{00}|$.) This observation is sufficient to permit the experimental extraction of $\theta_{+-}$. By measuring the rates for the 6 decays, one can construct the two triangles, as in Fig. 4. (In this figure, the $\tilde{A}$'s are related to the $\bar{A}$'s by a rotation.) Thus, up to a discrete ambiguity (since one or both triangles may be flipped upside-down), this determines $\theta_{+-}$. With this knowledge the angle $\alpha$ can be extracted by measuring CP violation in $B_d^0(t) \to \pi^+\pi^-$. Therefore, even in the presence of penguins, $\alpha$ can be obtained cleanly by using the above isospin analysis.

*1.4. Clean CP Violation without Tagging or Time Dependence*

All the examples given so far have required both tagging and time-dependent measurements in order to cleanly extract CKM phases. Experimentally, these are quite difficult. Is it possible to obtain clean weak phase information without tagging and time dependence? The answer to this question is *YES*.

One suggestion [9] is to use the decay $B^\pm \to D_{CP}^0 K^\pm$, i.e. to look for an asymmetry between $\Gamma(B^+ \to D_{CP}^0 K^+)$ and $\Gamma(B^- \to D_{CP}^0 K^-)$. Here, $D_{CP}^0$ is a $D^0$ or $\overline{D^0}$ which is identified in a CP-eigenstate mode (e.g. $\pi^+\pi^-$, $K^+K^-$, ...). Because charged $B$-mesons are involved, there is no mixing, so that neither tagging nor time dependence is needed.

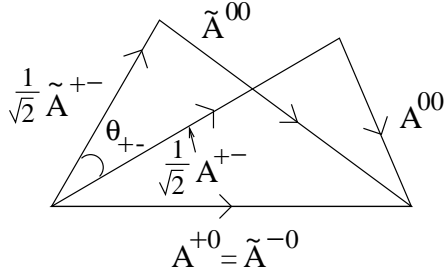

Figure 4: Isospin triangles in $B \to \pi\pi$.

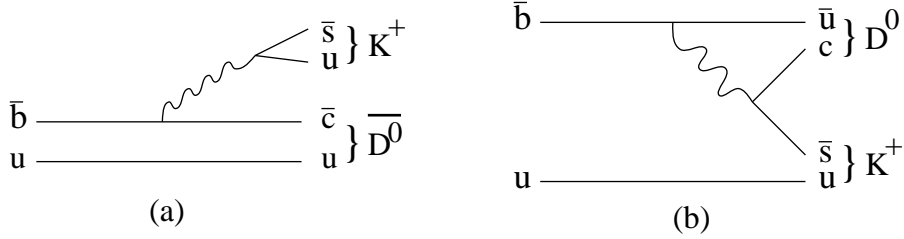

Figure 5: Diagrams contributing to (a) $B^+ \to \overline{D^0} K^+$ and (b) $B^+ \to D^0 K^+$.

A CP asymmetry in this decay would be a signal of direct CP violation, which requires two weak decay amplitudes. This is satisfied since the state $D^0_{CP}$ is in fact a linear combination of $D^0$ and $\overline{D^0}$:

$$D^0_{CP} = \frac{1}{\sqrt{2}}(D^0 + \overline{D^0}). \tag{7}$$

Thus, the two amplitudes come from the individual decays $B^+ \to D^0 K^+$ and $B^+ \to \overline{D^0} K^+$ (Fig. 5). These two amplitudes can be written

$$\begin{aligned} A(B^+ \to \overline{D^0} K^+) &= |A_1| e^{i\delta_1} , \\ A(B^+ \to D^0 K^+) &= |A_2| e^{i\gamma} e^{i\delta_1} , \end{aligned} \tag{8}$$

where $\delta_{1,2}$ are the strong phases of the two decays and $\gamma$ is the weak phase in $B^+ \to D^0 K^+$.

As discussed previously, simply measuring the direct CP violation in $B^\pm \to D^0_{CP} K^\pm$ does not yield the CKM angle $\gamma$ cleanly – the CP asymmetry is proportional to $\sin\gamma \sin(\delta_1 - \delta_2)$, and the strong phases are unknown. However, one can nevertheless extract $\gamma$ by using the triangle relations which follow from Eq. (7) above:

$$\begin{aligned} \sqrt{2} A(B^+ \to D^0_{CP} K^+) &= A(B^+ \to D^0 K^+) + A(B^+ \to \overline{D^0} K^+) , \\ \sqrt{2} A(B^- \to D^0_{CP} K^-) &= A(B^- \to \overline{D^0} K^-) + A(B^- \to D^0 K^-) . \end{aligned} \tag{9}$$

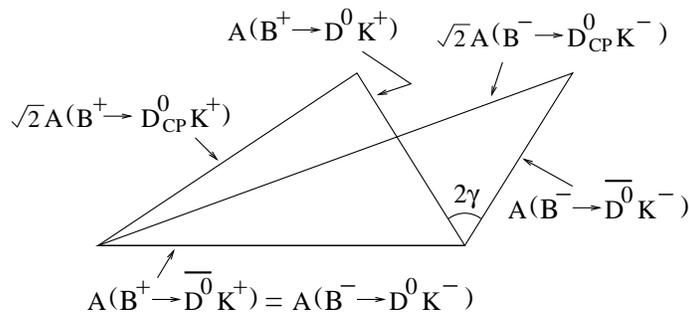

Figure 6: Triangles describing $B \to DK$.

Since there is only one amplitude which contributes to $B^+ \to D^0 K^+$ [Eq. (8)], there can be no CP-violating rate asymmetry between this decay and its CP-conjugate. The same holds for $B^+ \to \overline{D^0} K^+$. Thus we have

$$|A(B^+ \to D^0 K^+)| = |A(B^- \to \overline{D^0} K^-)|,$$
$$|A(B^+ \to \overline{D^0} K^+)| = |A(B^- \to D^0 K^-)|. \quad (10)$$

Note that there is a relative phase $2\gamma$ between $A(B^+ \to D^0 K^+)$ and $A(B^- \to \overline{D^0} K^-)$ [this follows from Eq. (8)]. Thus, the two triangles of Eq. (9) above have one side in common, and a second side of the same length. As to the third side, due to the possibility of CP violation, we expect that

$$|A(B^+ \to D^0_{CP} K^+)| \neq |A(B^- \to D^0_{CP} K^-)|. \quad (11)$$

The two triangles are shown in Fig. 6. From this figure one sees that, by measuring the 6 rates, it is possible to cleanly extract the angle $\gamma$, even though only direct CP violation is involved. There remains a discrete ambiguity [$\gamma \leftrightarrow (\delta_1 - \delta_2)$] due to the possibility of reflection of one of the triangles. Note that, even if there is no CP violation (i.e. if $\delta_1 = \delta_2$), $\gamma$ can still be obtained. Since neither tagging nor time-dependent measurements are necessary, this measurement can be done at a symmetric $B$-factory.

There is one possible problem with this technique. Although the decay $B^+ \to \overline{D^0} K^+$ is color-allowed, the decay $B^+ \to D^0 K^+$ is color-suppressed. This means that we expect the branching ratios for these two decays to be

$$BR(B^+ \to \overline{D^0} K^+) \sim 2 \times 10^{-4},$$
$$BR(B^+ \to D^0 K^+) \lesssim O(10^{-5}). \quad (12)$$

This means that the triangles are probably quite thin, which could make the extraction of $\gamma$ more difficult experimentally. As we will see in the next section, there are other techniques, based on flavor SU(3) symmetry, which allow the extraction of the weak and strong phases without tagging or time-dependence.

## 2. SU(3) RELATIONS AMONG AMPLITUDES

So far we have seen that the extraction of clean CKM phase information from direct CP violation is in general made impossible by our inability to calculate the strong phase shifts. Mixing-induced, or indirect, CP violation measurements are hampered by difficult time-dependent measurements and tagging, and at times by penguin pollution. However, we have seen that the use of SU(2) isospin symmetry enables us to disentangle these effects. Finally, while the use of $B \to DK$ decays provides a way to obtain clean CP violation information without tagging, only the angle $\gamma$ can be extracted this way and the triangle that needs to be constructed is expected to be very thin.

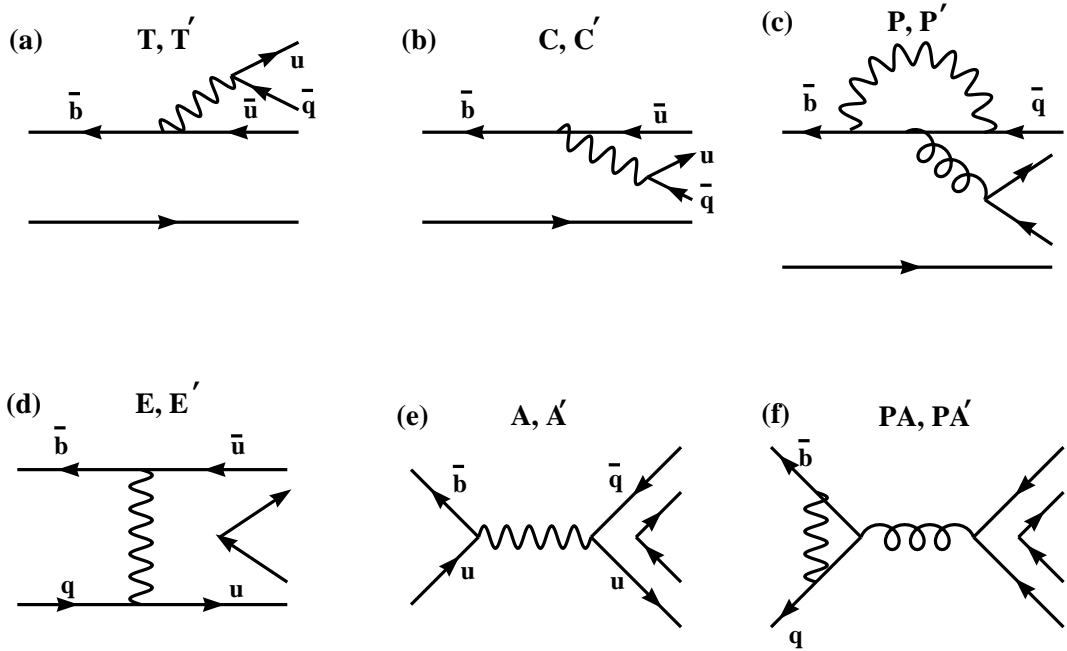

Figure 7: Diagrams describing decays of $B$ mesons to pairs of light pseudoscalar mesons. Here $\bar{q} = \bar{d}$ for unprimed amplitudes and $\bar{s}$ for primed amplitudes. (a) "Tree" (color-favored) amplitude $T$ or $T'$; (b) "Color-suppressed" amplitude $C$ or $C'$; (c) "Penguin" amplitude $P$ or $P'$ (we do not show intermediate quarks and gluons); (d) "Exchange" amplitude $E$ or $E'$; (e) "Annihilation" amplitude $A$ or $A'$; (f) "Penguin annihilation" amplitude $PA$ or $PA'$.

The successful application of isospin symmetry in the $B \to \pi\pi$ analysis leads quite naturally to the question, "what does flavor SU(3) imply?". The answer is "quite a lot!" [10]-[15]. We will see that, together with a few simple approximations, SU(3) symmetry allows us to obtain all of the CKM weak phases and all of the strong phase shifts from *time-independent* measurements alone.

In going from SU(2) to SU(3) the number of Goldstone bosons increases from 3 (the $\pi$'s) to 8 with the addition of $K, \overline{K}, K^+, K^-$ and the $\eta$ (we ignore the $\eta$ from now on because of its limited experimental usefulness). Following the conventions in Refs. [10, 13], we take the $u$, $d$, and $s$ quark to transform as a triplet of flavor SU(3), and the $-\bar{u}$, $\bar{d}$, and $\bar{s}$ to transform as an antitriplet. Thus the $\pi$-mesons and kaons form part of an octet and are defined as $\pi^+ \equiv u\bar{d}$, $\pi^0 \equiv (d\bar{d} - u\bar{u})/\sqrt{2}$, $\pi^- \equiv -d\bar{u}$,

$K^+ \equiv u\bar{s}$, $K^0 \equiv d\bar{s}$, $\bar{K}^0 \equiv s\bar{d}$ and $K^- \equiv -s\bar{u}$. The $B$ mesons, which are in the triplet or anti-triplet representation, are taken to be $B^+ \equiv \bar{b}u$, $B^0 \equiv \bar{b}d$, $B_s \equiv \bar{b}s$, $B^- \equiv -b\bar{u}$, $\overline{B}^0 \equiv b\bar{d}$ and $\overline{B}_s \equiv b\bar{s}$.

Consider all the decays of $B$ mesons to pairs of light pseudoscalar mesons $\pi\pi$, $\pi K$ and $K\bar{K}$. The amplitudes for these decays can be expressed in terms of the following diagrams (see Fig. 7): a "tree" amplitude $T$ or $T'$, a "color-suppressed" amplitude $C$ or $C'$, a "penguin" amplitude $P$ or $P'$, an "exchange" amplitude $E$ or $E'$, an "annihilation" amplitude $A$ or $A'$, and a "penguin annihilation" amplitude $PA$ or $PA'$. Here an unprimed amplitude stands for a strangeness-preserving decay, while a primed contribution stands for a strangeness-changing decay. As noted in Refs. [10, 13], this set of amplitudes is over-complete. The physical processes of interest involve only five distinct linear combinations of these six terms.

Now comes one of the main points. The diagrams denoted by $E$, $A$ and $PA$ can be ignored relative to the other diagrams. The reasons are as follows. First, the diagrams $E$ and $A$ are helicity suppressed by $(m_{u,d,s}/m_B)$ since the $B$ mesons are pseudoscalars. Second, annihilation and exchange processes, such as those represented by $E$, $A$, $PA$, are directly proportional to a factor of the $B$-meson wave function at the origin. Thus these diagrams are suppressed by a factor of $(f_B/m_B) \lesssim 0.05$ relative to diagrams $T$, $C$ and $P$ (and similarly for their primed counterparts). This suppression should remain valid unless hadronization and rescattering effects are important. Such rescatterings could be responsible for certain decays of charmed particles, but should be less important for the higher-energy $B$ decays.

Neglecting the contributions of the above diagrams, we are left with the 6 diagrams $T$, $T'$, $C$, $C'$, $P$ and $P'$. These six complex parameters determine the 13 allowed $B$ decays to states with pions and kaons, as listed in Table 1. This table is derived by expressing the $B$ into pseudoscalar decay as graphs in terms of their quark level contributions, keeping track of minus signs and $\sqrt{2}$ factors in going from quarks to mesons. The primed and unprimed diagrams are not independent, but are related by CKM matrix elements. In particular, $T'/T = C'/C = r_u$, where $r_u \equiv V_{us}/V_{ud} \approx 0.23$. Assuming that the penguin amplitudes are dominated by the top quark loop, one has $P'/P = r_t$, with $r_t \equiv V_{ts}/V_{td}$. We therefore have 13 decays described by 3 independent graphs, implying that there are 10 relations among the amplitudes. These can be expressed in terms of 6 amplitude equalities, 3 triangle relations, and one quadrangle relation.

The three independent triangle relations and one quadrangle relation are

$$(T + C) = (C - P) + (T + P) , \tag{13}$$

$$(T + C) = (C' - P')/r_u + (T' + P')/r_u , \tag{14}$$

$$(T + C) = (T' + C' + P')/r_u - (P')/r_u , \tag{15}$$

$$(T' + P') - (P') = r_u(T + P) - r_u(P) . \tag{16}$$

For example, by using Table 1 we can rewrite the relation in Eq. (13) in terms of decay amplitudes as:

$$\sqrt{2}A(B^+ \to \pi^+\pi^0) = \sqrt{2}A(B^0 \to \pi^0\pi^0) + A(B^0 \to \pi^+\pi^-) . \tag{17}$$

Table 1: The 13 decay amplitudes in terms of the 8 graphical combinations. The $\sqrt{2}(B^+ \to \pi^+\pi^0)$ in the $-(T+C)$ column means that $A(B^+\to\pi^+\pi^0)=-(T+C)/\sqrt{2}$, and similarly for other entries. Processes in the same column can be related by an amplitude equality, e.g. the amplitudes for $B^+\to K^+\overline{K}^0$ and $B^0\to K^0\overline{K}^0$ are equal.

| $-(T+C)$ | $-(C-P)$ | $-(T+P)$ | $(P)$ |
|---|---|---|---|
| $\sqrt{2}(B^+ \to \pi^+\pi^0)$ | $\sqrt{2}(B^0 \to \pi^0\pi^0)$ | $B^0 \to \pi^+\pi^-$ | $B^+ \to K^+\overline{K}^0$ |
| | $\sqrt{2}(B_s \to \pi^0\overline{K}^0)$ | $B_s \to \pi^+K^-$ | $B^0 \to K^0\overline{K}^0$ |
| $-(T'+C'+P')$ | $-(C'-P')$ | $-(T'+P')$ | $(P')$ |
| $\sqrt{2}(B^+ \to \pi^0 K^+)$ | $\sqrt{2}(B^0 \to \pi^0 K^0)$ | $B^0 \to \pi^- K^+$ | $B^+ \to \pi^+ K^0$ |
| | | $B_s \to K^- K^+$ | $B_s \to K^0\overline{K}^0$ |

We have chosen to express this relation using $B^0$ and $B^+$ mesons only. However, by using the amplitude equalities from Table 1, we could equally have written the right side of the above relation in terms of $B_s$.

## 3. MEASURING WEAK AND STRONG PHASES

### 3.1. Triangles, triangles, triangles

The surprising result [15] is that the three triangle relations allow us to *completely* solve for the magnitudes and phases of the amplitudes $T, C, P$. In addition we will have enough independent determinations of the same quantities to be able to test our two assumptions, namely SU(3) symmetry and the neglect of the $E, A, PA$ diagrams.

Since the amplitude for $B \to \pi^+\pi^0$ decay, given by $-(T+C)/\sqrt{2}$, is pure $\Delta I = 3/2$, the diagram $(T+C)$ has only one term, which we denote by $A_{I=2}e^{i\phi_2}e^{i\delta_2}$. Thus, for example, the triangle relation given in Eq. (13) becomes

$$A_{I=2}e^{i\phi_2}e^{i\delta_2} = (A_C e^{i\phi_C}e^{i\delta_C} - A_P e^{i\phi_P}e^{i\delta_P}) + (A_T e^{i\phi_T}e^{i\delta_T} + A_P e^{i\phi_P}e^{i\delta_P}) \quad , \qquad (18)$$

and similarly for the other relations. As before, the $\phi_i$ are the weak phases and the $\delta_i$ are the strong phases. The $\delta_i$ are chosen such that the quantities $A_{I=2}$, $A_T$, $A_{T'}$, $A_C$, $A_{C'}$, $A_P$ and $A_{P'}$ are real and positive (only relative strong phase differences are physically meaningful). SU(3) symmetry implies that the strong phases for the primed and unprimed graphs are equivalent. Working within the Wolfenstein approximation of the CKM matrix, it is easy to see that the weak phases of the various amplitudes are: $\phi_2 = \phi_T = \phi_{T'} = \phi_C = \phi_{C'} = \gamma$, $\phi_P = -\beta$, and $\phi_{P'} = \pi$ (up to corrections of order $\lambda^2 \approx 0.05$). Also, $A_{T'}/r_u = A_T$ and $A_{C'}/r_u = A_C$. Finally, multiplying through on both sides by $\exp(-i\gamma - i\delta_2)$, the 3 triangle relations become

$$\begin{align}
A_{I=2} &= (A_C e^{i\Delta_C} + A_P e^{i\alpha}e^{i\Delta_P}) + (A_T e^{i\Delta_T} - A_P e^{i\alpha}e^{i\Delta_P}), & (19) \\
A_{I=2} &= (A_C e^{i\Delta_C} + A_{P'} e^{-i\gamma}e^{i\Delta_P}/r_u) + (A_T e^{i\Delta_T} - A_{P'} e^{-i\gamma}e^{i\Delta_P}/r_u), & (20) \\
A_{I=2} &= (A_T e^{i\Delta_T} + A_C e^{i\Delta_C} - A_{P'} e^{-i\gamma}e^{i\Delta_P}/r_u) + A_{P'} e^{-i\gamma}e^{i\Delta_P}/r_u, & (21)
\end{align}$$

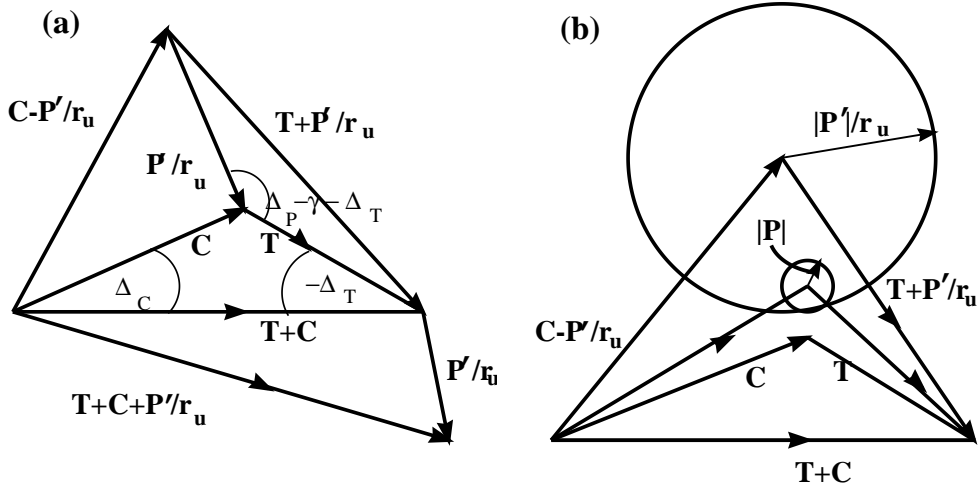

Figure 8: Triangle relations used to obtain weak phases and strong final-state phase shift differences. (a) Relation based on Eqs. (22) (upper triangle) and (21) (lower triangle). (b) Relation based on Eqs. (19) (lower triangle with small circle about its vertex) and (20) (upper triangle with large circle about its vertex). The relation based on (19) and (21) follows an almost identical construction.

where we have defined $\Delta_i \equiv \delta_i - \delta_2$.

Consider first the two triangle relations in Eqs. (20) and (21). These relations define two triangles which share a common base. Each triangle is determined up to a two-fold ambiguity, since it can be reflected about its base. Implicit in these two triangle relations is the relation

$$A_{I=2} = |T + C| = A_T e^{i\Delta_T} + A_C e^{i\Delta_C} \quad . \tag{22}$$

Thus both of these triangles also share a common subtriangle with sides $T + C$, $C$ and $T$ as shown in Fig. 8. The key point is this: the subtriangle is completely determined, up to a four-fold ambiguity, by the two triangles in Eqs. (20) and (21). This is because both the magnitude and relative direction of $P'/r_u$ are completely determined by constructing the triangle in Eq. (21). Therefore the point where the vectors $C$ and $T$ meet is given by drawing the vector $P'/r_u$ from the vertex opposite the base [see Fig. 8]. (A similar construction would have given the same point if we had used the vector $T + P'/r_u$ instead of $P'/r_u$.) Thus by measuring the five rates for

$B^0 \to \pi^0 K^0$ (giving $|C - P'/r_u|$),

$B^0 \to \pi^- K^+$ (giving $|T + P'/r_u|$),

$B^+ \to \pi^0 K^+$ (giving $|T + C + P'/r_u|$),

$B^+ \to \pi^+ K^0$ (giving $|P'/r_u|$), and

$B^+ \to \pi^+ \pi^0$ (giving $|T + C| = A_{I=2}$, i.e. the triangle's base),

we can determine $\Delta_P - \gamma$, $|T|$ and $|C|$, up to a two-fold ambiguity and $\Delta_C$ and $\Delta_T$ up to a four-fold ambiguity. As we will discuss later, these discrete ambiguities can be at least partially removed through the knowledge of the relative magnitudes of $|P|$, $|C|$, $|T|$ and $|P'|$, and through independent measurements of the amplitudes and the strong and weak phases.

If we also measure the rates for the CP-conjugate processes of the above decays, we can get more information. These CP-conjugate decays obey similar triangle relations to those in Eqs. (20) and (21). However, recall that under CP conjugation, the weak phases change sign, but strong phases do not. Thus we can perform an identical analysis with the CP-conjugate processes, giving us another, independent determination of $|T|$, $|C|$, $\Delta_C$ and $\Delta_T$. But, instead of $\Delta_P - \gamma$, this time we get $\Delta_P + \gamma$. Thus we obtain $\Delta_P$ and $\gamma$ separately. Note that it is not, in fact, necessary to measure all 5 CP-conjugate processes. The rate for $B^- \to \pi^- \pi^0$ is the same as that for $B^+ \to \pi^+ \pi^0$, since they involve a single weak phase and a single strong phase. Similarly, the rates for $B^+ \to \pi^+ K^0$ and $B^- \to \pi^- K^0$ are equal. Therefore, in order to extract $\gamma$, in addition to the above 5 rates, we need only measure $\overline{B}^0 \to \pi^0 \overline{K}^0$ (giving $|\bar{C} - \bar{P}'/r_u|$), $\overline{B}^0 \to \pi^+ K^-$ (giving $|\bar{T} + \bar{P}'/r_u|$), and $B^- \to \pi^0 K^-$ (giving $|\bar{T} + \bar{C} + \bar{P}'/r_u|$). To sum up, by measuring the above 8 rates, the following quantities can be obtained: the weak phase $\gamma$, the strong phase differences $\Delta_T$, $\Delta_C$ and $\Delta_P$, and the magnitudes of the different amplitudes $|T|$, $|C|$ and $|P'|$.

Note that the two triangles given by the relations in Eqs. (19) and (20) share a common base with each other and also with the sub-triangle in Eq. (22) (which still holds). The same is true for the two triangles constructed using the triangle relations in Eqs. (19) and (21). Unlike the first two-triangle construction, however, the shape of the sub-triangle is not yet fixed. Nevertheless, the point where the vectors $C$ and $T$ meet can still be determined by measuring the additional decays represented by $P$, $P'$, or $|T + P'/r_u|$. A detailed explanation of these two constructions can be found in Ref. [15]. The point is that by measuring 7 rates we can extract $\Delta_P + \alpha$, $\Delta_P - \gamma$, $\Delta_C$, and $\Delta_T$, up to an eight-fold ambiguity, and $|T|$ and $|C|$ up to a four-fold ambiguity. Through the two quantities $\Delta_P + \alpha$ and $\Delta_P - \gamma$, we can then determine the weak phase $\beta$ (using $\beta = \pi - \alpha - \gamma$), up to discrete ambiguities. As in the first two-triangle construction, all rates are time-independent. What is surprising, perhaps, about this particular construction is that *it is not even necessary to measure the CP-conjugate rates in order to obtain $\beta$*. The reason is that SU(3) flavor symmetry implies the equality of the strong final-state phases of two different amplitudes, in this case $P$ and $P'$. Subtracting the (strong plus weak) phase of one amplitude from the other then determines a weak phase. Usually, in a given process, without measuring the charge-conjugate rate one can only measure the sum of a weak and a strong phase.

If the CP-conjugate rates are also measured, we can obtain $\Delta_P$, $\alpha$, and $\gamma$ separately. This provides another, independent determination of $|T|$, $|C|$, $\Delta_C$ and $\Delta_T$. As in the first construction, no observation of CP violation is necessary to make such measurements. Again, it is not necessary to measure all the CP-conjugate rates – only four can be different from their counterparts.

*3.2. Testing our assumptions*

The three constructions use $B$ decays to $\pi\pi$, $\pi K$ and $K\bar{K}$ final states. At present, the decays $B^0 \to \pi^+\pi^-$ and/or $\pi^-K^+$ have been observed, but the two final states cannot be distinguished [16]. The combined branching ratio is about $2 \times 10^{-5}$. Assuming equal rates for $\pi^+\pi^-$ and $\pi^-K^+$, which seems likely, the amplitudes $|T|$ and $|P'|$ should be about the same size. On the other hand, the amplitude $|C|$ is expected to be about a factor of 5 smaller: the amplitudes $|T|$ and $|C|$ are basically the same as $|a_1|$ and $|a_2|$, respectively, introduced in Ref. [17], for which the values $|a_1| = 1.11$ and $|a_2| = 0.21$ have been found [18]. The ratio $|P/T|$ has also been estimated to be small, $\lesssim 0.20$ [7]. Therefore all the decays used in these constructions should have branching ratios of the order of $10^{-5}$, with the exception of $B \to K\bar{K}$ ($P$) and $B^0 \to \pi^0\pi^0$ [$\sim (C-P)$], which are probably an order of magnitude smaller.

The knowledge that the amplitudes obey the hierarchy $|P|, |C| < |T| < |P'/r_u|$ will also help in reducing discrete ambiguities. For example, in the first two-triangle construction [Fig. 8], we noted in the discussion following Eq. (22) that the subtriangle can be determined up to a four-fold ambiguity. However, two of these four solutions imply that $|C|$ and $|T|$ are both of order $|P'/r_u|$, which violates the above hierarchy. Thus the four-fold ambiguity in the determination of the subtriangle is reduced to a two-fold ambiguity, and the discrete ambiguities in the determination of subsequent quantities such as $\Delta_P - \gamma$, $\Delta_C$, etc., are likewise reduced. The ambiguities in the other two constructions can be partially removed in a similar way.

All three two-triangle constructions described above rely on two assumptions. The first is that the diagrams $A$, $E$ and $PA$ (and their primed counterparts) can be neglected. This can be tested experimentally. The decays $B^0 \to K^+K^-$ and $B_s \to \pi^+\pi^-$ can occur only through the diagrams $E$ and $PA$, and $E'$ and $PA'$, respectively. Therefore, if the above assumption is correct, the rates for these two decays should be much smaller than the rates for the decays in Table 1.

The second assumption is that of an unbroken SU(3) symmetry. We know, however, that SU(3) is in fact broken in nature. Assuming factorization, SU(3)-breaking effects can be taken into account by including the meson decay constants $f_\pi$ and $f_K$ in the relations between $B \to \pi\pi$ decays and $B \to \pi K$ decays [12]. In other words, the factor $r_u$ which appears in two of the triangle relations should be multiplied by $f_K/f_\pi \approx 1.2$. One way to test whether this properly accounts for all SU(3)-breaking effects is through the rate equalities in Table 1. Even if it turns out that $f_K/f_\pi$ does not take into account all SU(3)-breaking effects, the large number of independent measurements is likely to help in reducing uncertainties due to SU(3) breaking. For example, note that, not counting the CP-conjugate processes, the last two constructions have six of their seven rates in common. This means that a measurement of only eight decay rates gives two independent measurements of $|T|$, $|C|$, $\Delta_C$, $\Delta_T$, $\Delta_P - \gamma$ and $\Delta_P + \alpha$. In fact, these eight rates already contain the five rates of the first construction [Fig. 8]. Thus we actually have three independent ways of arriving at $|T|$, $|C|$, $\Delta_C$, $\Delta_T$ and $\Delta_P - \gamma$. Including also the CP-conjugate processes, we have a total of 13 $B$-decay rate measurements which give us six independent ways to measure $|T|$, $|C|$, $\Delta_C$ and $\Delta_T$, five ways to measure $\Delta_P$, three independent ways to measure $\gamma$, and two ways to measure $\alpha$. (If time-dependent measurements are possible, there are additional independent ways to measure $\alpha$.) The point is that the three two-triangle constructions include many ways to measure the same quantity. This redundancy provides a powerful way to test the validity of our SU(3) analysis and reduces the discrete ambiguities in the determination of the various quantities.

## 4. Conclusions

The measurement of the angles $\alpha$, $\beta$ and $\gamma$ of the unitarity triangle will be a crucial test of the SM picture of CP violation. Although there are many decays which are likely to exhibit CP violation, very few provide clean information on $\alpha$, $\beta$ and $\gamma$, due to the presence of incalculable strong phase shifts and penguin diagrams. We have reviewed the ways in which the CKM angles can be obtained cleanly. The angles $\alpha$, $\beta$ and $\gamma$ can be extracted from the time-dependent measurements of the rates for $\overset{\scriptscriptstyle(-)}{B_d} \to \pi^+\pi^-$, $\overset{\scriptscriptstyle(-)}{B_d} \to \Psi K_S$ and $\overset{\scriptscriptstyle(-)}{B_s} \to D_s^+ K^-, D_s^- K^+$, respectively. In the case of $\overset{\scriptscriptstyle(-)}{B_d} \to \pi^+\pi^-$, it may be necessary to use an isospin analysis to remove the unwanted penguin contributions. The angle $\gamma$ can also be obtained by looking at the decays $B^\pm \to D^0_{CP} K^\pm$. The advantage of this method is that neither tagging nor time dependence is necessary; the disadvantage is that the triangles used in this analysis are likely to be quite thin, which would make a precise determination of $\gamma$ difficult.

We have also described in some detail the recent developments which provide a prescription for the measurement of all relevant quantities: weak and strong phases, and the sizes of the contributing diagrams. This analysis uses SU(3) flavour symmetry along with the important dynamical assumption that exchange and annihilation diagrams can be neglected. This method relies on several triangle relations which hold under these assumptions. Like $B \to DK$ decays, neither time-dependent measurements nor tagging are required. This analysis can therefore be carried out at a symmetric $B$-factory. Unlike $B \to DK$, however, the branching ratios for most of the processes involved are expected to be $O(10^{-5})$, so that the sides of the triangles are all roughly the same size. This method also provides enough redundancy to test the consistency of the assumptions.

## Acknowledgements

We thank M. Gronau, B. Kayser, R.D. Peccei and J.L. Rosner for collaborations and helpful conversations. We also thank J.-R. Cudell and K. Dienes for the opportunity to speak at this very well-organized MRST conference. This work was supported in part by the N.S.E.R.C. of Canada and les Fonds F.C.A.R. du Québec.